\documentclass[prx,aps,twocolumn]{revtex4-1}

\usepackage[pdftex]{graphicx}
\usepackage{amsbsy,amssymb,amsmath,bm,mathtools}

\usepackage{xspace}
\usepackage{bm}% bold math
\usepackage{color}

\usepackage{url}
\usepackage[section]{placeins}
\usepackage[colorlinks=true,linkcolor=blue,citecolor=blue]{hyperref}

\begin{document}

\preprint{APS/123-QED}

\setlength{\abovecaptionskip}{-0pt}

\title{Anomalous and Anisotropic Nonlinear Susceptibility in the Proximate Kitaev Magnet $\alpha$-RuCl$_3$}

\author{Ludwig Holleis$^1$}
\author{Joseph Prestigiacomo$^2$}
\author{Zhijie Fan$^1$}
\author{Satoshi Nishimoto$^3$}
\author{Mike Osofsky$^2$}
\author{Gia-Wei Chern$^1$}
\author{Jeroen van den Brink$^{3,4}$}
\author{B.S. Shivaram$^1$}
\email{https://orcid.org/0000-0002-6792-7463}

\affiliation{$^1$Department of Physics, University of Virginia, Charlottesville, VA. 22904}
\affiliation{$^2$Naval Research Laboratory, 4440 Overlook Drive, Washington D.C, USA}
\affiliation{$^3$Department of Physics, Technical University Dresden, 01069 Dresden, Germany}
\affiliation{$^4$Institute for Theoretical Solid State Physics, IFW Dresden, 01069 Dresden, Germany}

\keywords{Quantum Spin Liquid, Nonlinear Susceptibility, Fractional Excitations, Majorana Fermions, Kitaev Magnet, Metamagnet}

\begin{abstract}
The leading order nonlinear (NL) susceptibility, $\chi_3$, in a paramagnet is negative and diverges as $T \rightarrow 0$.  This divergence is destroyed when spins correlate and the NL response provides unique insights into magnetic order.  Dimensionality, exchange interaction, and preponderance of quantum effects all imprint their signatures in the NL magnetic response. Here, we study the NL susceptibilities in the proximate Kitaev magnet $\alpha$-RuCl$_3$ which differs from the expected antiferromagnetic behavior. For $T< T_c$ = 7.5 K and field $B$ in the ab-plane, we obtain contrasting NL responses in low ($<$ 2 ${T}$) and high field regions. For low fields the NL behavior is dominated by a quadratic response (positive $\chi_2$), which shows a rapid rise below $T_c$. This large $\chi_2 >0$ implies a broken sublattice symmetry of magnetic order at low temperatures. Classical Monte Carlo (CMC) simulations in the standard ${K-H-\Gamma}$ model secure such a quadratic ${B}$ dependence of ${M}$, only for ${T}$ $\approx$ ${T}_c$  with ${\chi}_2$ being zero as ${T}$ $\rightarrow$ 0. It is also zero for all temperatures in exact diagonalization calculations. On the other hand, we find an exclusive cubic term (${\chi}_3$) describes the high field NL behavior well.   ${\chi}_3$ is large and positive both below and above ${T}_c$ crossing zero only for ${T}$ $>$ 50 K. In contrast, for $B$~$\parallel$~c-axis, no separate low/high field behaviors is measured and only a much smaller ${\chi}_3$ is apparent.
\end{abstract}

\maketitle

\section{Introduction}
Since the demonstration by Kitaev~\cite{KitaevAnnPhys2006} of the existence of a quantum spin liquid state in two dimensions through exact calculations on a honeycomb lattice, there has been an intense experimental search for its realization in real material systems.  Aided through insights provided by Jackeli and Khaliullin~\cite{JackeliPRL2009} to engineer (bond dependent) exchange interactions, the materials search has identified several promising systems with prominent attention given thus far to two candidates, the iridium oxides and ruthenium chloride \cite{HermannsRev2017, SavaryRev2017}.  Despite the presence of a very large exchange energy (of the order of 100 K) these systems do not order down to comparatively low temperatures. When they do order magnetically, the features observed in both microscopic probes such as neutron scattering and macroscopic measurements such as magnetometry and thermal response are highly unusual and contrast dramatically with what is expected from conventional magnets.  In their thermal transport~\cite{KasaharaNature2018}, thermodynamic~\cite{DoNatPhys2017}, and microwave response~\cite{WellmPRB2018}, they provide tell-tale signs as to the presence of fractionalized excitations sought after in Kitaev magnets.

In the proximate Kitaev spin liquid candidate, $\alpha$-RuCl$_3$ neutron scattering experiments show a low temperature magnetic excitation spectrum consisting of sharp spin wave peaks and a continuum associated with fractional excitations~\cite{BanerjeeNatMat2016, DoNatPhys2017}. A magnetic transition that sets in at 7.5 K where the spins assume a zig-zag chain pattern located in the ab-plane with two neighboring chains being antiferromagnetically aligned~\cite{BanerjeeScience2017} is also established.  Raman scattering studies also reveal unconventional magnetic excitations with a broad continuum whose temperature dependence is apparent over a large scale compared to the magnetic ordering temperature~\cite{SandilandsPRL2015, YipingWangNPJQ2020}.  The linear susceptibility shows a discontinuity in slope~\cite{SearsPRB2015} at ${T}_N$ = 7.5 K and exhibits substantial in-plane anisotropy that persists in the normal state~\cite{LampenKelleyPRB2018}. The high temperature ($T$ $>$ 150 K) susceptibility is convincingly Curie like, however, there is an extended intermediate ``Kitaev paramagnetic" region~\cite{DoNatPhys2017} beyond ${T}_c$. The out of plane anisotropies are also significant: the susceptibility parallel to the c-axis is nearly an order of magnitude smaller with only a minor signature at the 7.5 K transition.  These magnetic signatures are in stark contrast to what is known in conventional 2D (insulating) antiferromagnets~\cite{NaritaJPSJ1996}. 

In this communication we report measurements of the nonlinear DC susceptibilities, ${\chi}_2$ and ${\chi}_3$, in $\alpha$-RuCl$_3$ and illustrate that they probe many key aspects of the proximate Kitaev spin liquid state as well as the zigzag antiferromagnetic phase. The equilibrium magnetization in any system can be written in the general form:
\begin{equation}
M = \chi_1 B + \chi_2 B^2 +\chi_3 B^3 + \cdots
\end{equation} 
where the coefficients represent the various order susceptibilities. In particular, the coefficient ${\chi}_2$ is non zero when time reversable symmetry is broken, while preserving lattice symmetry as for instance \cite{Bitla2011} in ferromagnets.

	The nonlinear susceptibility ${\chi}_3$ in a classical paramagnet is negative at all temperatures and diverges~\cite{MorinAndSchmitt1981} as ${T}$ $\rightarrow$ 0 while ${\chi}_2$ is non existent due to time reversal symmetry.  The negative divergence in ${\chi}_3$ can be interrupted however if the system develops long range magnetic order. We illustrate in fig. 1(a-c) the three known types of characteristic behavior of both the linear (${\chi}_1$) and the nonlinear (${\chi}_2$ and ${\chi}_3$) susceptibilities for typical ferromagnets, bipartite antiferromagnets and spin glasses, respectively~\cite{Bitla2011, GingrasPRL1997, RamirezPRL1990, SuzukiPTP1977}. In all three cases, significant non zero ${\chi}_3$ is found only in the vicinity of the critical temperature.  It is worth noting that ${\chi}_3$ assumes only negative values at $T> T_c$ for all three types of magnets and tends to zero rather quickly as ${T}$ increases.  The second order susceptibility, ${\chi}_2$, is less studied and a non zero ${\chi}_2$ is possible only when time reversal symmetry is either explicitly or spontaneously broken, as in ferromagnets.  Moreover, even for bipartite anti ferromagnets, an effective time reversal symmetry due to the sublattice symmetry results in a vanishing ${\chi}_2$ even in the ordered state below ${T}_c$; see Fig.1(B). \\

In $\alpha$-RuCl$_3$ we find in contrast that ${\chi}_3$ assumes significantly positive values over an extended temperature range above the ordering temperature.  While there is a divergence of ${\chi}_3$ at ${T}_c$ as might be expected, its value remains significantly positive down to the lowest temperatures.  We also find a quadratic field dependence of ${M}$, giving rise to a significantly positive ${\chi}_2$ which exists only in the ordered state and has a large non-zero value even as ${T}$ $\rightarrow$ 0.  To our knowledge such behavior of ${\chi}_3$ and ${\chi}_2$ have not been observed before.  Further, this unexpected quadratic contribution is highly anisotropic and found only when the magnetic field is perpendicular to the high symmetry axis (c-axis).  \\  

\section{Results and Discussion}

\subsection{Experimental results}
In figure 2, we show the measured magnetization isotherms for ${B}$ $\parallel$ a-axis (i.e. $\phi=0^0$ as per the definition adopted in \cite{LampenKelleyPRB2018}) plotted in a manner that facilitates the extraction of the nonlinear susceptibilities. Equation~(1) above defines the susceptibility parameters and motivates plots such as Fig.~2. It is apparent from the top nine panels in Fig.~2 that the slope of the lines for B $\parallel$ a-axis which are well defined and close to zero at high ${T}$, turn significantly positive when lowering ${T}$, and increase in magnitude as the ordering temperature ${T}_c$ = 7.5 K is approached. Below ${T}_c$, the response breaks into two distinct regions, with a crossover threshold field of ${B}$* $\approx$ 2 ${T}$. The nonlinear response below B* is considerably larger than at the high field end, 3\,T~$<B<$~5\,T.  Confining ourselves to the quadratic term only, as per Eq. 1 in this low field region, we obtain values of ${\chi}_2$ as shown in Fig. 3b.  The values of ${\chi}_2$ are positive and large at the lowest temperatures and decrease monotonically towards ${T}_c$, where it rapidly drops to zero. It is important to note that a non-zero ${\chi}_2$ is only possible in systems in which time-reversal symmetry is broken\cite{LaiCTP2018}.  In a strict bi-partite antiferromagnet, however, time reversal symmetry is not broken. The fits described above also yield  ${\chi}_1$ via the intercept which can be used as a consistency check on the linear susceptibility values obtained at constant low field through temperature sweeps. The absence of any feature in ${\chi}_1$ in the vicinity of 14 K alludes to the high quality of the sample measured.  \\

\begin{figure*}
\includegraphics[width=0.9\textwidth]{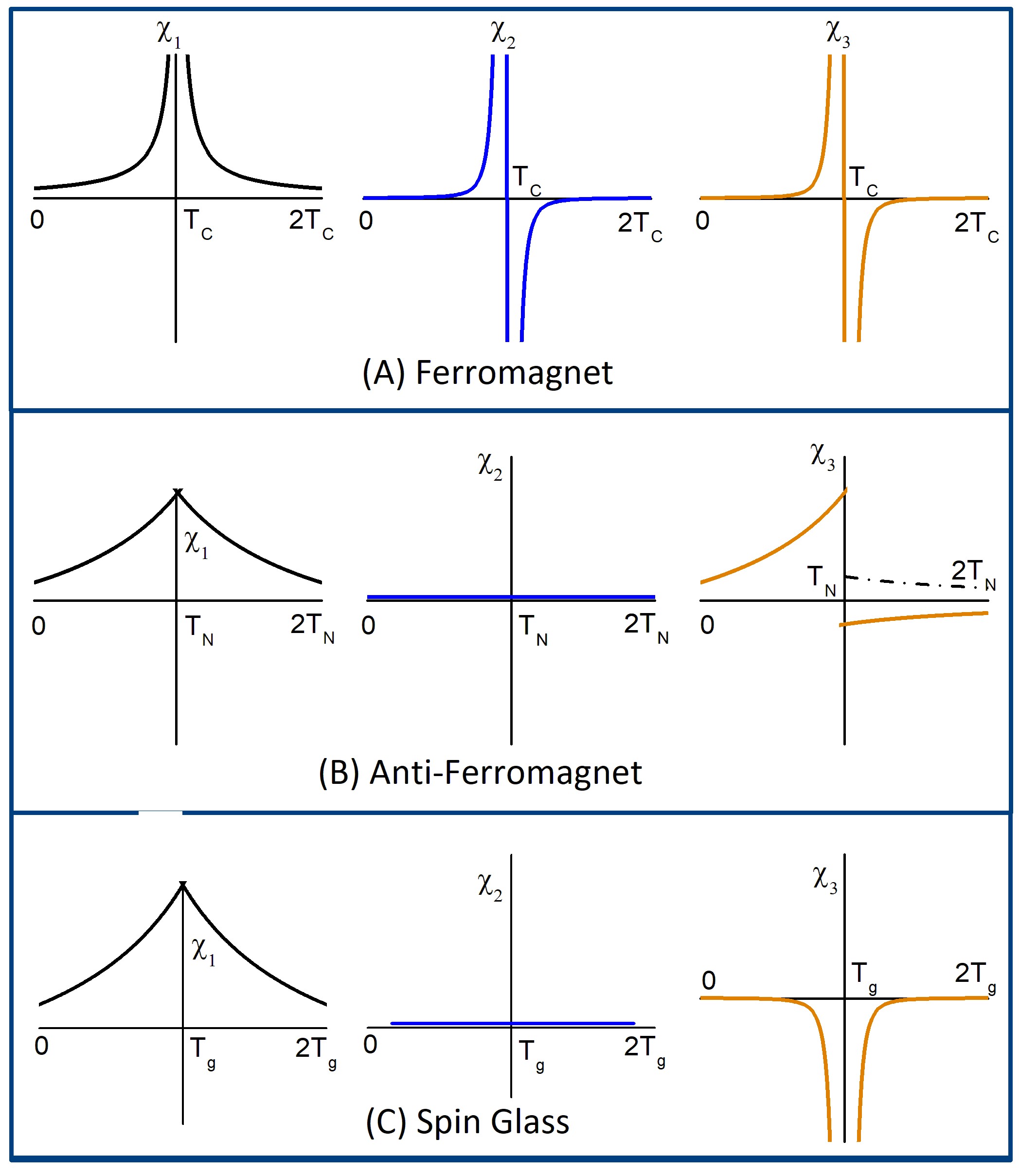}
\vspace{0.5cm}
\caption{\label{fig1}{\textbf{Schematic of the known qualitative behaviors of the linear (${\chi}_1$) and nonlinear (${\chi}_2$ and ${\chi}_3$) magnetic susceptibilities in different flavors of magnetic systems:}  In types A and C, the nonlinear susceptibility, ${\chi}_3$, is negative above the characteristic temperature and has a sharp discontinuity at a characteristic temperature (${T}_c$ or ${T}_g$). In type B, the sign of ${\chi}_3$ in the paramagnetic regime depends on the coordination number\cite{FujikiPTP1980}. Below this temperature it is observed to rapidly approach zero as ${T}$ $\rightarrow$ 0 for all three phases. ${\chi}_2$ on the other hand exhibits such a discontinuity only for type A (e.g. ferromagnet) where time reversal symmetry is broken\cite{Bitla2011}. In comparison, in the other two cases it is zero.}}
\end{figure*}

\begin{figure*}
\includegraphics[width=0.9\textwidth]{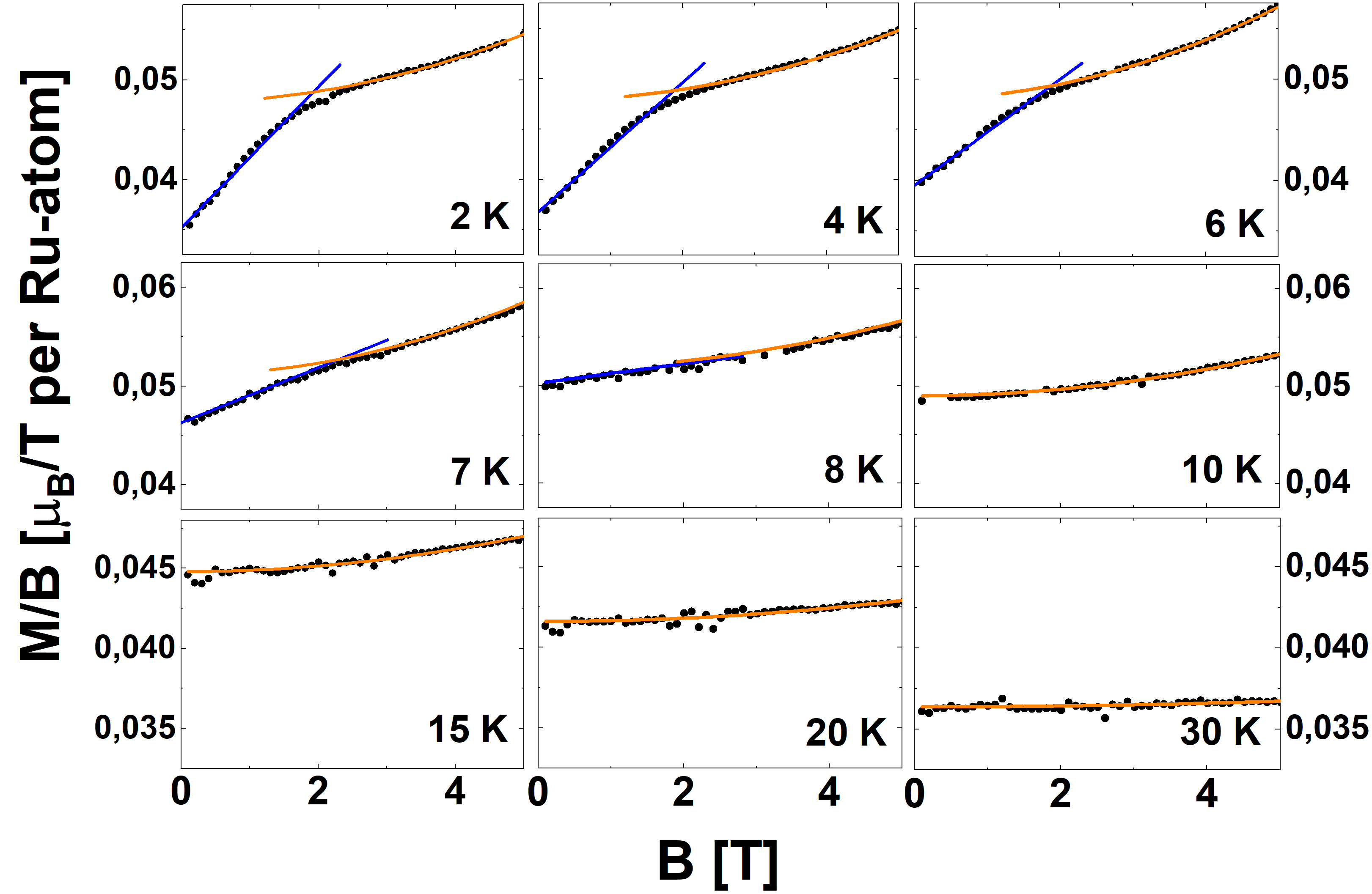}
\includegraphics[width=0.9\textwidth]{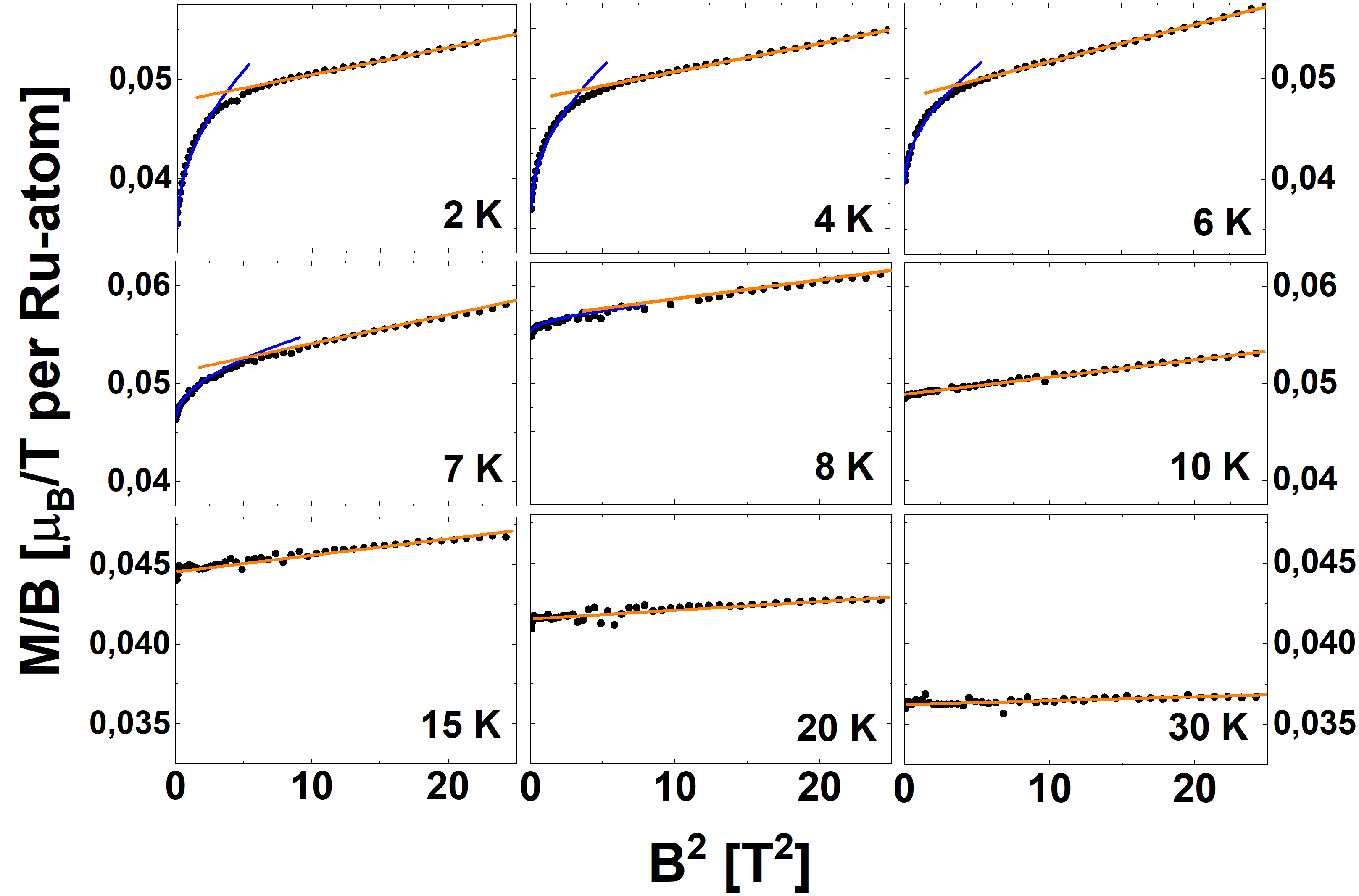}
\vspace{0.3cm}
\caption{\label{fig2}\textbf{Magnetization isotherms:} The top set of nine panels show the ratio of the measured magnetization, ${M}$, to the magnetic field, ${B}$, in $\alpha$-RuCl$_3$ (with ${B}$ up to 5 T $\parallel$ a-axis) plotted against ${B}$. Such a plot provides the quadratic nonlinear susceptibility, ${\chi}_2$, via the slope of the blue straight lines on the low field side. Since at the high field end (${B}$ $>$ 3 T) the response it better fit with a cubic term ${\chi}_3$, we show in the nine panels on the bottom a similar plot but with ${B}^2$ on the abscissa and the fits in orange straight lines. It is clear from these panels that ${\chi}_3$ when ${B}$ $\parallel$ a-axis is positive over a wide temperature range, while ${\chi}_2$ vanishes for ${T}$ $>$ ${T}_N$. }
\end{figure*}

\begin{figure*}
\includegraphics[width=1.0\textwidth]{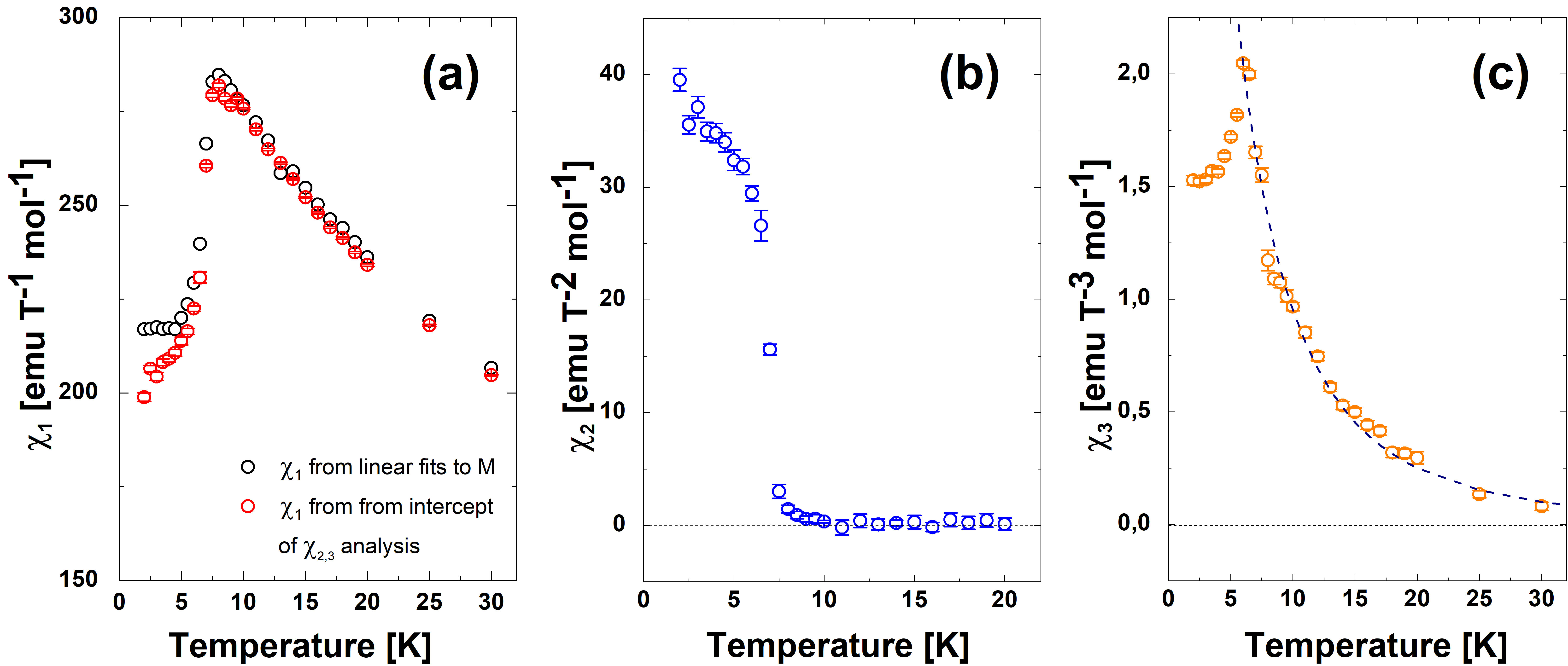}
\vspace{0.7cm}
\caption{\label{fig3} \textbf{Linear and higher order susceptibilities:} Panel (a) shows the  linear susceptibility obtained through temperature sweep (black dots) as well as those obtained from the intercepts of the magnetization isotherms of the ${\chi}_2$ and ${\chi}_3$ analysis such as in Fig.2 (red dots). Panel (b) shows ${\chi}_2$ extracted from the low field quadratic response, from plots such as in fig.2, top nine panels. Similarly, (c) shows the temperature dependence of ${\chi}_3$ obtained from plots such as those in fig. 2, bottom nine panels. A sharp "lambda"-like anomaly is apparent for ${\chi}_3$ obtained from the high field (3 T - 5 T) response when ${B}$ $\parallel$ a-axis. The error bars capture the coefficient of determination of the linear fits such as shown in fig 2.}
\end{figure*}

Another significant feature notable in the nine panels on the top in Fig.~2 is the presence of a clear upward curvature, in the 3 T - 5 T region, particularly in the 6 K and 10 K isotherms.  This curvature implies the presence of a cubic term which can be extracted by plotting M/B against B$^2$ on the x-axis as shown in the bottom nine panels of Fig.~2. The values of ${\chi}_3$ obtained by fitting the linear region in such plots with B between 3 T and 5 T below ${T}_c$ and over the entire field range (0T - 5T) above ${T}_c$, are shown in Fig.~3, panel~c. The high linearity of these fits illustrates the absence of a ${\chi}_2$ term in these regions of the phase diagram. Performing such separate fits in the two regions is the most natural way to analyze our results. Our approach of seperate fits is further motivated by signs of a crossover transition at $\approx$ 2 ${T}$ in  neutron \cite{SearsPRB2015, BanerjeeNPJQ2018}, microwave response\cite{WellmPRB2018} and differential susceptibility\cite{KubotaPRB2015} experiments.\\

 As can be seen in  Fig.~3c ${\chi}_3$ is significantly positive near ${T}_c$ and decreases in magnitude as the temperature is increased.  A power law fit as shown in the figure provides a reasonable empirical description of the behavior of ${\chi}_3$ in the (Kitaev) paramagnetic region.  Experimentally however, a crossover to negative ${\chi}_3$ occurs above 50 K (a temperature of the order of the Kitaev-exchange\cite{DoNatPhys2017}).  Below ${T}_c$ the values of the nonlinear susceptibility are obtained through linear fits in two separate regions as explained above. The high field region yields values for ${\chi}_3$ that exhibit a sharp peak at ${T}$ = 6.5 K just below the ordering temperature.  They remain positive down to the lowest temperature of 2 K with no evidence of a downturn towards zero.  While a negative cubic power law dependence is expected from the Curie-law for the third order susceptibility in paramagnets \cite{MorinAndSchmitt1981}, such a positive behavior has not been reported to our knowledge.\\

We have also measured the nonlinear susceptibility for magnetic fields perpendicular to the ab-plane. Representative magnetization isotherms plotted similar to those in Fig.~2 are shown in supplementary information, Fig.~S2. The extracted values of ${\chi}_3$ are small and positive and increase monotonically as the temperature is lowered as seen in Fig.~S3. More importantly, as evident in Fig.~S2, the nonlinear response is well described by a single cubic term and there is no evidence for a large quadratic response as found  for ${B}$ $\parallel$ a-axis. Given that for ${B}$ $\parallel$ a-axis the deviation from a linear behavior of the magnetization due to a quadratic contribution is roughly ten times larger than that arising from the cubic term, such behavior if present for ${B}$ $\parallel$ c-axis - would be easily seen within the resolution of the MPMS SQUID measurements.  Considering the anisotropy factor of approximately eight as observed experimentally in linear susceptibility between the ab-plane and the c-axis and assuming that this is entirely due to the g-factor, one can predict an anisotropy in ${\chi}_3$ of $\approx$ 8$^3$ = 512 in reasonable agreement with our experimental results. We also present in the supplementary section, Fig.~S4, preliminary data for the in plane nonlinear response when ${B}$ $\perp$ a-axis.  While the behavior observed is qualitatively the same as in Fig.~3, there are quantitative differences.  Further characterization of this in-plane anisotropy will form part of a separate comprehensive study.\\ 

\subsection{Comparison to Traditional Nonlinear Susceptibility Responses}
Measurements of the equilibrium third order susceptibility, although rare, have been performed in 2D magnets, frustrated systems, spin glasses and strongly correlated itinerant metamagnets, that is materials which show a rapid rise of magnetization at a critical field ${B}_c$ \cite{ShivaramPRB2014}.  In a bipartite antiferromagnet such as FeCl$_2$ \cite{Kushauer1995}, or even in the classic 2D magnets, (C$_2$H$_5$NH$_3$)$_2$CuCl$_4$ \cite{NaritaJPSJ1996}, the DC nonlinear susceptibility, ${\chi}_3$ exhibits a "lambda" like anomaly, just below ${T}_c$ as expected by theory.  From below ${T}_c$ the standard response of ${\chi}_3$  in these systems is a rapid rise to a large positive value at ${T}_c$, above which it drops with the sign being set by the coordination number \cite{FujikiPTP1980}.  For instance, in the paramagnetic phase of the Kagome system SCGO \cite{Schiffer1996} a large positive value of ${\chi}_3$ is seen above ${T}_c$, but it reaches close to zero within 2${T}_c$.  In many strongly correlated itinerant metamagnets \cite{RamirezPRL1992, Kitagawa1996} in which an order-parameter develops, ${\chi}_3$ peaks at the ordering temperature, and decreases rapidly as ${T}$ $\rightarrow$ 0.  Nonlinear susceptibility measurements in such systems typically probe higher order correlations and place strong constraints on the ground state \cite{RamirezPRL1994, BauerPRB2006}.  These latter systems are three dimensional electronically but can exhibit a strong magnetic anisotropy due to the ${g}$-factor. Nevertheless, in these system, for all directions it is sufficient to include only the cubic term and the possibility of a dual response with a quadratic term has almost never been discussed \cite{ZheludevPRL1997}.  Thus, the features reported above in $\alpha$-RuCl$_3$ are in contrast to much of what is known about nonlinear susceptibilities in magnets. The unmistakable presence of a large ${T}$ $\rightarrow$0 quadratic term makes $\alpha$-RuCl$_3$ a unique 2D quantum antiferromagnet.    \\

\subsection{Comparison to Simulations of the $J_1$-$J_3$-K-$\Gamma$ Model}
In order to understand the nature of the nonlinear susceptibility in $\alpha$-RuCl$_3$ it is possible to consider several approaches based on different model Hamiltonians employed thus far \cite{LampenKelleyPRB2018,WinterPhysRevLett2018}. Most of these approaches start with the Kitaev-Heisenberg model appended with various choices of off-diagonal terms \cite{WinterJPCM2017, JanssenPRB2017}.  The correct choice of the Hamiltonian is still very much a matter of debate with the sign of the Kitaev term or even the necessity of the Kitaev term being in question \cite{Yao-Dong2019, EvanWilson2021, YoungKim2021}. We focus on this model and apply two separate calculational tools to study the nonlinear susceptibilities for $\alpha$-RuCl$_3$: (a) we use classical Monte Carlo (CMC) simulations with the model Hamiltonian

\begin{eqnarray}
\mathcal{H} &=&  \sum_{{\langle ij \rangle}_\gamma} [J_1 \mathbf{S}_i \cdot \mathbf{ S}_j + K S_i^\gamma  S_j^\gamma + \Gamma (S_i^\alpha S_j^\beta + S_i^\beta S_j^\alpha)] \nonumber \\
& & +\sum_{\langle \langle \langle ij \rangle \rangle \rangle} J_3 \mathbf{S}_i \cdot \mathbf{ S}_j - g \mu_B \mathbf B \cdot  \sum_i \mathbf{S}_i
\end{eqnarray}

\begin{figure*}
\includegraphics[width=0.6\textwidth]{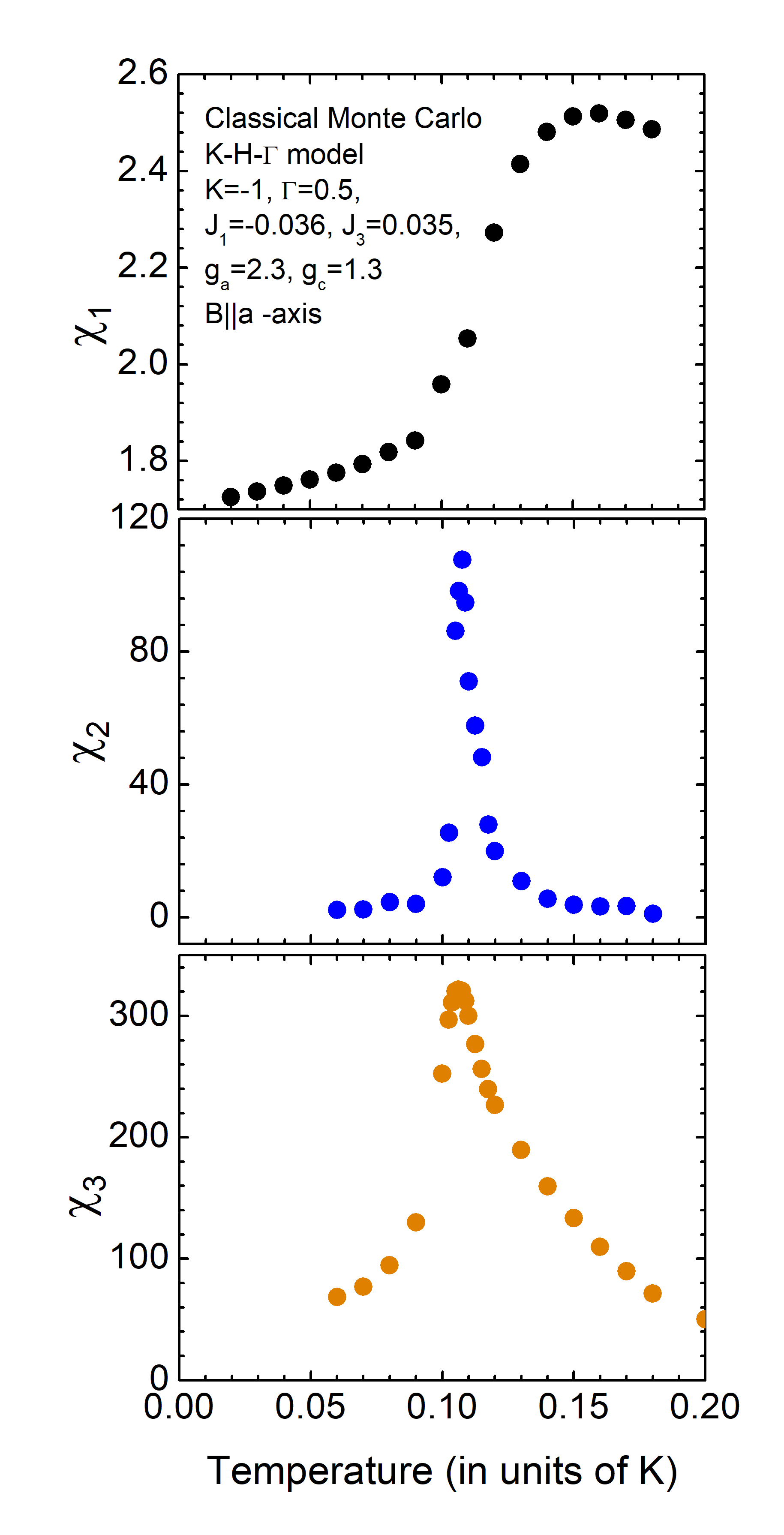}
\vspace{-0.4cm}
\caption{\label{fig4}\textbf{CMC simulations:} Shows the temperature dependence of the linear and nonlinear susceptibilities obtained in Monte Carlo calculations, for ${B}$ $\parallel$ a-axis. ${\chi}_2$ rises rapidly to positive values at ${T}_N$ reaches a maximum and drops sharply again below ${T}_N$. This latter feature is in sharp contrast to the experimental results where the large positive ${\chi}_2$ persists to the lowest temperatures measured. Similar results in contrast to experiment are obtained for the behavior of ${\chi}_3$. It also attains a positive value below ${T}$ $\approx$ 0.11 (in units of K$_1$), reaches a peak value near ${T}_N$ and rapidly decreases at lower temperatures. Note that y-axis scale is arbitrary.}
\end{figure*}

This so-called ${J}_1-{J}_3-{K}-{\Gamma}$ model is considered one of the the most successful efforts in modeling $\alpha$-RuCl$_3$~\cite{WinterPhysRevLett2018}.
(b) we employ exact diagonalization methods with slight adjustment on parameters as in \cite{YadavSR2016}.  For (a) although quantum fluctuations are expected to be strong for such spin-1/2 systems, the fact that the magnet develops a long range antiferromagnetic order justifies the classical spin approximation, as a first step to understanding the magnetic properties of the ordered phase. For (b) the choice is based on the recognition that it reproduces the magnetization isotherms to high fields very well.\\

\begin{figure*}
\includegraphics[width=0.5\textwidth]{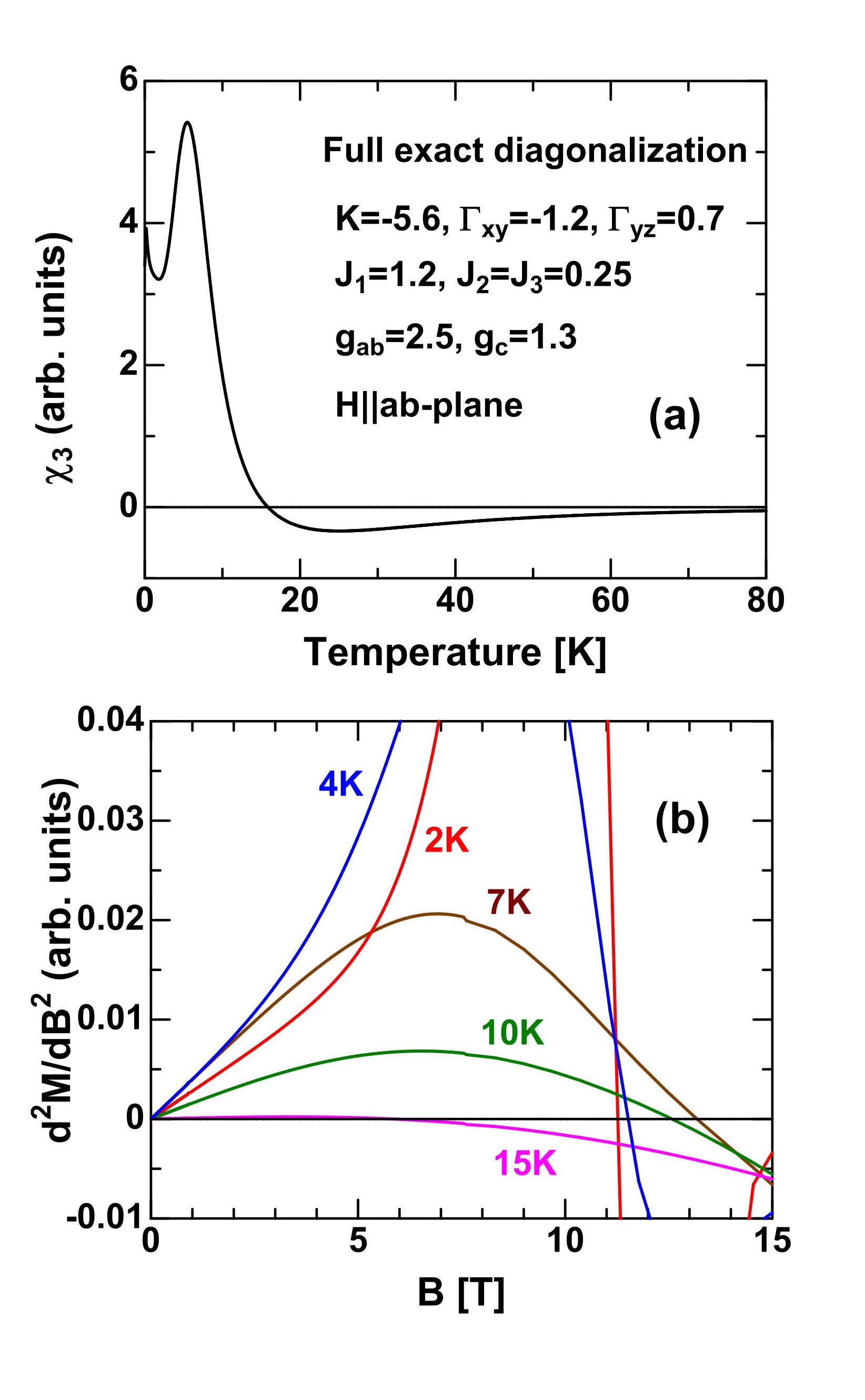}
\vspace{-0.4 cm}
\caption{\label{fig5} \textbf{Exact Diagonlization Calculations:} Shows the nonlinear susceptibility ${\chi}_3$ in calculations employing exact diagonalization, for ${B}$ in the ab plane.  The results obtained are identical whether ${B}$ $\parallel$ a-axis or ${B}$ $\perp$ a-axis.  ${\chi}_3$ rises rapidly to positive values at ${T}_N$ reaches a maximum and saturates to a large positive value at ${T}$ = 0.  However, the zero crossing occurs within 2${T}_N$ whereas the experimental zero crossing occurs at a much higher temperature.  The bottom panel shows the second derivative, ${d^2M/dB^2}$, at various temperatures.  This derivative passes zero at all temperatures - thus implying the absence of a ${B}^2$-term in the magnetization contrary to experimental findings.}
\end{figure*}

With the CMC simulations we reproduce the phase transition to the so called zig zag order at ${T}_c$ $\approx$ 0.11 K where K is the dominant Kitaev exchange constant, with the other parameter values normalized to K taken as, ${\Gamma}$ = 0.5, ${J}_1$ = 0.036, ${J}_3$ = 0.035, ${g}_a$ = ${g}_b$ = 2.3 and ${g}_c$ = 1.3 \cite{WinterNature2017, KimPRB2016}. Our calculations of the linear susceptibility ${\chi}_1$ based on the Monte Carlo data, as shown in Fig.~4(a), shows a broad bump near the critical temperature, a feature that is qualitatively consistent with the experiment, Fig.~3(a).  However, in stark contrast to the experimental data, our 
simulations find a positive and significant ${\chi}_2$ only in the vicinity of ${T}_c$, Fig.~4(b).  Our finding of a vanishing ${\chi}_2$ at very low temperatures in the zig-zag phase is in fact consistent with an emergent time reversal symmetry of the low ${T}$ ordered state.  To see this emergent symmetry we first note that although there are three equivalent propagation directions for the zig zag order, it has been shown experimentally and theoretically that the ground state of $\alpha$-RuCl$_3$ is a single ${Q}$ zig zag.  Spins in such single ${Q}$ zig zag are collinear, with opposite spins sitting on alternating zig zag chains of the honey comb lattice.  The total magnetization of such collinear bipartite antiferromagnetic order is the sum of the two sublattices: $\mathbf M = (M_A + M_B) \,\hat{\mathbf n}$, where $\hat{\mathbf n}$  is the collinear spin direction, and $\langle M_B \rangle = - \langle M_A \rangle$. Their contributions to the second order magnetic susceptibility d
\begin{equation}
	\chi_2 \sim \langle M^3 \rangle = \langle M_A^3 \rangle + \langle M_B^3\rangle + 3 \langle M_A^2  M_B \rangle + 3 \langle M_A M_B^2 \rangle  
\end{equation}
cancel each other due to the sublattice symmetry $\langle M_A^3 \rangle = - \langle M_B^3\rangle$ and $\langle M_A^2 M_B \rangle = - \langle M_B M_A^2 \rangle$.  The persistence of the quadratic ${\chi}_2$ down to the lowest temperatures in our experiments thus highlights the unusual nature of the zig zag order.  This intriguing discrepancy could be due to non trivial quantum fluctuations in the ordered state.  Another possible scenario is that the non zero ${\chi}_2$ results from a highly inhomogenous multidomain zig zag order at low temperatures.  Indeed, at the interfaces between different zig zag domains, the two sub lattice collinear spins argument given above no longer applies and a non zero ${\chi}_2$ could result from the complex non-collinear structure at the domain boundaries.  Our experiments also imply that the low temperature ${\chi}_2$ vanishes at the critical field ${B*}$ $\approx$ 2 T.  Interestingly, this critical field ${B*}$ has recently been attributed to the so called Q flop transition identified both in neutron diffraction \cite{SearsPRB2015, BanerjeeNPJQ2018} and terahertz
 spectroscopy  \cite{WuPRB2018, OzelPRB2019}.  Since the presence of the magnetic field in the ab-plane breaks the equivalence of the three zig zag orientations, the Q-flop transition describes a repopulation of zig zag domains in which two energetically unfavorable zig zag domains are replaced by the third one.  Such a realignment of the zig zag domains also significantly reduces the non collinear spins residing on the interfaces of different zig zag order, thus giving rise to a reduced ${\chi}_2$.  Detailed numerical simulations of the Q-flop transition and its effects on ${\chi}_2$ will be left for a further computational study. The breaking of the sublattice-symmetry, which leads to a nonzero ${\chi}_2$ in the ground-state, could also come at the dynamical level, for example, due to non trivial quantum fluctuations with non-collinear high-order spin-correlations in the ordered state \cite{RamirezPRL1992}.  Other possibilities such as a stacking of the single-Q collinear zigzag order along the c direction that breaks the sublattice-symmetry cannot be ruled out either. Compared with CMC simulations another intriguing result from our experiments is the persistence of large ${\chi}_3$ in both high and low temperature regimes.  The values of ${\chi}_3$ from CMC simulations approach zero very rapidly above ${T}_c$ in contrast to the experimental large and positive values that persist for temperatures significantly greater than ${T}_c$.  The low temperature discrepancy of ${\chi}_3$ can be attributed to the absence of quantum fluctuations in CMC. \\
\vspace{2.0 mm}

However, the importance of such fluctuations is borne out in similar calculations utilizing quantum chemistry methods discussed below.  They are also seen in quantum Monte Carlo simulations in the pure Kitaev limit by Kamiya \cite{KamiyaAPS2019} which show a persistent positive ${\chi}_3$ down to the lowest temperatures. In quantum chemistry methods we use exact diagonalization (ED) of a closely related Hamiltonian employed in \cite{YadavSR2016} which yields accurate results for the magnetization isotherms in good agreement with experimental results at the high field end. The calculated magnetization isotherms in this approach plotted in a manner similar to experiments, are well fit with a single straight line (see Fig.~S9) implying that only ${\chi}_3$ contributes. The values of ${\chi}_3$ extracted from such calculations is large and positive even at ${T}$ = 0, Fig.~5. Also shown in Fig.~5 is the behavior of the calculated derivative d$^2$M/dB$^2$ which displays a zero intercept for all temperatures. This implies that the dual slope response, which we attribute to the presence of the complex multi-domain zigzag order or other sublattice-symmetry breaking mechanisms, is not found in the ED calculations. It is worth noting that the non-zero value of ${\chi}_3$ for ${T}$ $\rightarrow$ 0 is also found in the pure Kitaev model with Quantum Monte Carlo calculations \cite{KamiyaAPS2019}. In such calculations it is possible to secure a crossover of ${\chi}_3$ at a higher temperature, however only through the antiferromagnetic Kitaev interaction. Many calculations \cite{WinterPhysRevLett2018} rule out an antiferromagnetic scenario, but our experimental results suggest not to exclude this possibility.\\

Moving forward, any viable model for $\alpha$-RuCl$_3$ must explain the quadratic contribution to the magnetization in the antiferromagnetic zigzag phase evident in our experiments. Further, it also has to account for the persistence of the large positive values of the third order susceptibility for temperatures well above ${T}_c$. It might be that the details of the material parameters will decide the magnitude and the temperature range over which a positive nonlinearity is stretched. 
\\

\subsection{Summary and Outlook}

In summary, we have presented measurements of the nonlinear susceptibility in the Kitaev magnet $\alpha$-RuCl$_3$. Most significantly, our work has uncovered an anomalous quadratic response of the magnetization to field which yields a large positive  ${\chi}_2$ in the ordered state as ${T}$ $\rightarrow$ 0. This behavior is absent when B $\parallel$ c-axis suggesting a strong 2D nature of the order parameter. This anisotropy as well as the measured anisotropy of ${\chi}_3$ both above and below ${T}_c$ can serve as future characterization tools for pinning down specific models for proximate Kitaev materials. In addition, our observation of extended positive behavior of ${\chi}_3$ up to 50 K is consistent with previous circumstantial evidence that Kitaev type behavior with associated excitations persist up to fairly large temperatures ${T}$ $\approx$ 60 K $>>$ ${T}_c$ \cite{SandilandsPRL2015, DoNatPhys2017}. The low field crossover i.e. the anomalous response with a quadratic term at low fields does not have to be confined to $\alpha$-RuCl$_3$ and the generality of its presence in Kitaev or other spin liquid compounds should be established. Furthermore, our results could place constraints on models which attempt to explain experimental observations in $\alpha$-RuCl$_3$ and similar compounds outside the realm of Kitaev physics  \cite{ModicArxiv2019, Yao-Dong2019}. Is the anomalous ${\chi}_2$ and the extended positive ${\chi}_3$ a natural consequence of  Kitaev type models when quantum fluctuations are correctly accounted for or is it a peculiarity of $\alpha$-RuCl$_3$ \cite{Dai2DMatls2020}? We note that the conditions to approach pure Kitaev without the nuisance of magnetic order experimentally are fairly easy to reach - the zig zag order is destroyed in $\alpha$-RuCl$_3$ at relatively low pressures of $\approx$ 1 GPa \cite{CuiPRB2018, WangPRB2018}.  Equally important to carry this work forward will be attempts to predict an in plane anisotropy of the nonlinear response in a quantitative manner followed by further detailed experimental work in this regard.

% Methods and Materials

\section{Methods}
\subsection{Materials and Experiments}
Our measurements were performed on high quality single crystals similar to those used for recent linear magnetometry measurements\cite{LampenKelleyPRB2018}. A Quantum Design Magnetic Property Measurement System SQUID magnetometer capable of reaching 5 T was employed to obtain magnetization isotherms in the temperature range 2 - 300 K. \\

\subsection{Simulations}
Classical Monte Carlo as well as quantum chemistry based exact diagonalization (ED) calculations of the linear and nonlinear magnetic response in the ${K-H-\Gamma}$ model\cite{WinterPhysRevLett2018}, were performed using available packages on a supercomputer.

% Acknowledgements
\medskip
\textbf{Acknowledgements} \par %delete if not applicable))
We acknowledge many useful conversations with and suggestions for experimental work from Steve Nagler, David Mandrus and Arnab Banerjee.  We are grateful to Jiaqiang Yan for the sample of single crystal $\alpha$-RuCl$_3$.  We acknowledge useful discussions and correspondence with Yoshitomo Kamiya, Joji Nasu, Nandini Trivedi, Collin Broholm and Mike Zhitomirsky. Z. Fan and G.-W. Chern are partially supported by the Center for Materials Theory as a part of the Computational Materials Science (CMS) program, funded by the US Department of Energy, Office of Science, Basic Energy Sciences, Materials Sciences and Engineering Division. The work of B.S. Shivaram was supported by NSF Award DMR-2016909. The work in Germany was supported by DFG through SFB 1143 project A05.\\

\medskip
\textbf{Author Contribution Statement}
Ludwig Holleis (LH) and Joseph Prestigiacomo collected the data.  LH and B.S. Shivaram (BSS) analysed the data and made the plots.  Zhije Fan and Gia-Wei Chern (GWC) performed the classical Monte Carlo simulations.  Satoshi Nishimoto and Jeroen van den Brink calculated the magnetization through exact diagonalization of the Hamiltonian.
 Michael Osofsky, BSS and GWC planned the research and developed the manuscript which was edited by all authors.

\medskip
\textbf{Competing Interests}
The authors declare no competing interests.

\medskip
\textbf{Data availability statement}
The datasets generated during and/or analysed during the current study are available from the corresponding author on reasonable request.

% References


\medskip


\begin{thebibliography}{10}

\bibitem{KitaevAnnPhys2006}
A.~Kitaev, Anyons in an exactly solved model and beyond, \newblock \emph{Ann. Phys.} \emph{321}, 2-111, \textbf{2006}.

\bibitem{JackeliPRL2009}
G.~Jackeli, and G.~Khaliullin, Mott insulators in the strong spin-orbit coupling limit: from Heisenberg to a quantum compass and Kitaev models, \newblock \emph{Phys. Rev. Lett.}, \emph{102}, 017205,\textbf{2009}.

\bibitem{HermannsRev2017}
J.~K. M.~Hermanns, I.~Kimchi, and J.Knolle, Physics of the Kitaev Model: Fractionalization, Dynamic Correlations, and Material Connections, \newblock \emph{Annu. Rev. Condens. Matter Phys.}, \emph{9} 17-33,\textbf{2018}.

\bibitem{SavaryRev2017}
L.~Savary, and L.~Balents, Quantum spin liquids: a review, \newblock \emph{Rep. Prog. Phys.}, \emph{80} 016502,\textbf{2017}.

\bibitem{KasaharaNature2018}
Y.~Kasahara et al., Majorana quantization and half-integer thermal quantum Hall effect in a Kitaev spin liquid, \newblock \emph{Nature}, \emph{559}, 227-231, \textbf{2018}.

\bibitem{DoNatPhys2017}
S.-H. Do et al., Majorana fermions in the Kitaev quantum spin system $\alpha$-RuCl$_3$,  \newblock \emph{Nat. Phys.} , \emph{13} 1079-1084, \textbf{2017}.

\bibitem{WellmPRB2018}
C. Wellm et al., Signatures of low-energy fractionalized excitations in $\alpha$-RuCl$_3$  from field-dependent microwave absorption, \newblock \emph{Phys. Rev. B}, \emph{98} 184408, \textbf{2018}.

\bibitem{BanerjeeNatMat2016}
A.~Banerjee et al., Proximate Kitaev quantum spin liquid behaviour in a honeycomb magnet, \newblock \emph{Nat. Mater.}, \emph{15} 733-740, \textbf{2016}.

\bibitem{BanerjeeScience2017}
A.~Banerjee et al., Neutron scattering in the proximate quantum spin liquid $\alpha$-RuCl$_3$, \newblock \emph{Science}, \emph{356}, 1055-1059,\textbf{2017}.

\bibitem{SandilandsPRL2015}Luke J. Sandilands, Yao Tian, Kemp W. Plumb, and Young-June Kim, and Kenneth S. Burch, Scattering continuum and possible fractionalized excitations in $\alpha$-RuCl$_3$, Phys. Rev. Lett., \emph{114}, 147201, \textbf{2015}.

\bibitem{YipingWangNPJQ2020}Yiping Wang et al., The range of non-Kitaev terms and fractional particles in $\alpha$-RuCl$_3$, npj Quantum Materials, \emph{5}, 14, \textbf{2020}.

\bibitem{SearsPRB2015}
J.~A. Sears et al., Magnetic order in $\alpha$-RuCl$_3$: A honeycomb-lattice quantum magnet with strong spin-orbit coupling, \newblock \emph{Phys. Rev. B} , \emph{91} 144420, \textbf{2015}.

\bibitem{LampenKelleyPRB2018}
P.~Lampen-Kelley et al., Anisotropic susceptibilities in the honeycomb Kitaev system $\alpha$-RuCl$_3$, \newblock \emph{Phys. Rev. B}, \emph{98} 100403(R),  \textbf{2018}.

\bibitem{NaritaJPSJ1996}
N.~Narita, I.~Yamada, Nonlinear Magnetic-Susceptibility of Two-Dimensional Magnets ( C$_n$H$_{2n+1}$NH$_3$)$_2$CuCl$_4$ with n=1, 2 and 3, \newblock \emph{J. Phys. Soc. Jpn.} \textbf{1996}, \emph{65} 4054.

\bibitem{Bitla2011}
Y.~Bitla, S.~N. Kaul, Mean-field treatment of nonlinear susceptibilities for a ferromagnet of arbitrary spin, \newblock \emph{Europhys. Lett.},  \emph{96} 37012, \textbf{2011}.

\bibitem{MorinAndSchmitt1981}
P.~Morin, D.~Schmitt, Third-order magnetic susceptibility as a new method for studying quadrupolar interactions in rare-earth compounds, \newblock \emph{Phys. Rev. B}, \emph{23} 5936-5949,  \textbf{1981}.

\bibitem{GingrasPRL1997}
M.~J.~P. Gingras et al., Static Critical Behavior of the Spin-Freezing Transition in the Geometrically Frustrated Pyrochlore Antiferromagnet ${\mathrm{Y}}_{2}{\mathrm{Mo}}_{2}{\mathrm{O}}_{7}$,  \newblock \emph{Phys. Rev. Lett.}, \emph{78} 947-950, \textbf{1997}.

\bibitem{RamirezPRL1990}
A.~P. Ramirez, G.~P. Espinosa, A.~S. Cooper, Strong frustration and dilution-enhanced order in a quasi-2D spin glass, \newblock \emph{Phys. Rev. Lett.}, \emph{64} 2070-2073 \textbf{1990}.

\bibitem{SuzukiPTP1977}
M.~Suzuki, Phenomenological Theory of Spin-Glasses and Some Rigorous Results,
\newblock \emph{Prog. Theor. Phys.}, \emph{58} 1151-1165, \textbf{1977}.

\bibitem{LaiCTP2018}
L.-Q. Lai, Z.~Li, Y.-B. Yu, Q.-H. Liu, Third-Order Magnetic Susceptibility of an Ideal Fermi Gas, \newblock \emph{Commun. Theor. Phys.}, \emph{70}, 619, \textbf{2018}.

\bibitem{BanerjeeNPJQ2018}
A.~Banerjee et al., Excitations in the field-induced quantum spin liquid state of $\alpha$-RuCl$_3$, \newblock \emph{npj Quantum Mater.}, \emph{3} 8, \textbf{2018}.

\bibitem{KubotaPRB2015}
Y.~Kubota, H.~Tanaka, T.~Ono, Y.~Narumi, K.~Kindo, Successive magnetic phase transitions in $\alpha$-RuCl$_3$: XY-like frustrated magnet on the honeycomb lattice, \newblock \emph{Phys. Rev. B}, \emph{91} 094422, \textbf{2015}.

\bibitem{ShivaramPRB2014}
See for~example: B.~S.~Shivaram, D. G.~Hinks, P.~Kumar, M. Andrade and M.B. Maple, Universality in the magnetic response of metamagnetic metals, \newblock \emph{Phys. Rev. B}, \emph{89} 241107(R),  \textbf{2014}.

\bibitem{Kushauer1995}
J.~Kushauer, W.~Kleemann, Critical behaviour of the linear and non-linear magnetic susceptibilities of FeCl$_2$, \newblock \emph{J. Phys.: Condens. Matter}, \emph{7} L1-L6, \textbf{1995}.

\bibitem{FujikiPTP1980}
S.~Fujiki, S.~Katsura, Nonlinear Susceptibility in the Spin Glass, \newblock \emph{Prog. Theor. Phys.}, \emph{64} 1130, \textbf{1981}.

\bibitem{Schiffer1996}
P.~Schiffer, A.~Ramirez, K.~N. Franklin, S.~W. Cheong, Interaction-Induced Spin Coplanarity in a Kagom\'e Magnet: SrCr ${}_{9\mathit{p}}$Ga ${}_{12\ensuremath{-}9\mathit{p}}$${\mathrm{O}}_{19}$, \newblock \emph{Phys. Rev. Lett.} , \emph{77} 2085-2088, \textbf{1995}.

\bibitem{RamirezPRL1992}
A.~P. Ramirez et al., Nonlinear susceptibility as a probe of tensor spin order in ${\mathrm{URu}}_{2}$${\mathrm{Si}}_{2}$ , \newblock \emph{Phys. Rev. Lett.}, \emph{68} 2680-2683, \textbf{1992}.

\bibitem{Kitagawa1996}
J.~Kitagawa et al., Third-Order Magnetic Susceptibility and Quadrupolar Order
Parameter of Kondo-Lattice Compound Ce$_3$Pd$_20$Ge$_6$, \newblock \emph{J. Phys. Soc. Jpn.}, \emph{69} 883-887, \textbf{2007}.

\bibitem{RamirezPRL1994}
A. P. Ramirez, ' P. Chandra, P. Coleman, Z. Fisk, J.L. Smith, and H. R. Ott, Nonlinear Susceptibility: A Direct Test of the Quadrupolar Kondo Effect in U${\mathrm{Be}}_{13}$, Phys. Rev. Lett., \emph{73}, 3018-3021, \textbf{1994}.

\bibitem{BauerPRB2006}E. D. Bauer et al., Nonlinear susceptibility: Evidence for antiferroquadrupolar fluctuations and a nonmagnetic ${\ensuremath{\Gamma}}_{1}$ ground state in the heavy fermion superconductor ${\mathrm{PrOs}}_{4}{\mathrm{Sb}}_{12}$., Phys.  Rev. \emph{B73}, 094511, \textbf{2006}. 

\bibitem{ZheludevPRL1997}
For an exception~see: A. Zheludev, S. Maslov, G. Shirane, Y. Sasago, N. Koide, and K. Uchinokura, Field-Induced Commensurate-Incommensurate Phase Transition in a Dzyaloshinskii-Moriya Spiral Antiferromagnet, Phys. Rev. Lett., \emph{78}, 4857-4860, (1997). 

\bibitem{WinterPhysRevLett2018}
S.~M. Winter, K.~Riedl, D.~Kaib, R.~Coldea, R.~Valenti, Probing $\alpha$-RuCl$_3$ Beyond Magnetic Order: Effects of Temperature and Magnetic Field, \newblock \emph{Phys. Rev. Lett.}, \emph{120} 077203,  \textbf{2018}.

\bibitem{WinterJPCM2017}Stephen M Winter et al., Models and materials for generalized Kitaev magnetism, J. Phys.: Condens. Matter, \emph{29}  493002, \textbf{2017}.

\bibitem{JanssenPRB2017}Lukas Janssen, Eric C. Andrade, and Matthias Vojta, Magnetization processes of zigzag states on the honeycomb lattice: Identifying spin models for $\alpha$-RuCl$_3$  and Na$_2$IrO$_3$, Phys. Rev. B, \emph{96}, 064430,\textbf{2017}.

\bibitem{Yao-Dong2019}
Y.~Z. Yao-Dong~Li, Xu~Yang, G.~Chen, Non-Kitaev spin liquids in Kitaev materials, \newblock \emph{Phys. Rev. B}, \emph{99} 205119, \textbf{2019}.

\bibitem{EvanWilson2021} Evan Wilson and Jason Haraldsen, Understanding the magnetic interactions of the zig-zag honeycomb lattice: Application to RuCl<sub>3</sub>, https://meetings.aps.org/Meeting/MAR21/Session/E39.11, {2021}.

\bibitem{YoungKim2021}Young-June Kim, Ferromagnetic Kitaev interactions and magnetic anisotropy in $\alpha$-RuCl$_3$, https://meetings.aps.org/Meeting/MAR21/Session/A54.1, {2021}.

\bibitem{YadavSR2016}
R.~Yadav et al., Kitaev exchange and field-induced quantum spin-liquid states in honeycomb $\alpha$-RuCl$_3$, \newblock \emph{Sci. Rep.}, \emph{6} 37925,  \textbf{2016}.

\bibitem{WinterNature2017}
S. M. Winter et al., Breakdown of magnons in a strongly spin-orbital coupled magnet, Nat. Commun. \emph{8}, 1152, \textbf{2017}.

\bibitem{KimPRB2016} H.-S. Kim and H.-Y. Kee, Crystal structure and magnetism in $\alpha$-RuCl$_3$: An ab initio study, Phys. Rev. \emph{B 93}, 155143, \textbf{2016}.

\bibitem{WuPRB2018}L. Wu et al., Field evolution of magnons in $\ensuremath{\alpha}\text{\ensuremath{-}}{\mathrm{RuCl}}_{3}$ by high-resolution polarized terahertz spectroscopy, Phys. Rev. B \emph{98}, 094425 \textbf{2018}. 

\bibitem{OzelPRB2019}I. O. Ozel et al., Magnetic field-dependent low-energy magnon dynamics in $\ensuremath{\alpha}--{\mathrm{RuCl}}_{3}$, Phys. Rev. B, \emph{100}, 085108 \textbf{2019}. 

\bibitem{KamiyaAPS2019}
Y.~Kamiya, J.~Yoshitake, Y.~Kato, J.~Nasu, Y.~Motome, Nonlinear magnetic susceptibility in the Kitaev model, https://meetings.aps.org/Meeting/MAR19/Session/A37.3, \textbf{2019}.

\bibitem{ModicArxiv2019}
K.~Modic, et al., \newblock \emph{arXiv:2005.04228} \textbf{2020}.

\bibitem{Dai2DMatls2020}Zhongwei Dai et al., Crystal structure reconstruction in the surface monolayer of the quantum spin liquid candidate $\alpha$-RuCl$_3$, 2D Materials, \emph{7}, 035004, \textbf{2020}.

\bibitem{CuiPRB2018}
Y.~Cui, J.et al., High-pressure magnetization and NMR studies of $\alpha$-RuCl$_3$, \newblock \emph{Phys. Rev. B}, \emph{96} 205147,  \textbf{2017}.

\bibitem{WangPRB2018}
Z.~Wang, et al., Pressure-induced melting of magnetic order and emergence of a new quantum state in $\alpha$-RuCl$_3$, \newblock \emph{Phys. Rev. B},  \emph{97} 245149, \textbf{2018}.


\end{thebibliography}
\end{document}